\documentclass{PoS}
\usepackage{psfrag,epsfig,graphicx,graphics,xcolor}
\usepackage{wrapfig}
\usepackage{amsmath}
\title{A model for high energy rho meson leptoproduction based on collinear factorization and dipole models}

\ShortTitle{High energy rho meson leptoproduction}

\author{\speaker{Adrien Besse}
\\
        LPT, Universit\'e Paris-Sud, CNRS, 91405, Orsay, France\\
        National Center for Nuclear Research (NCBJ), Warsaw, Poland\\
        E-mail: \email{adrien.besse@th.u-psud.fr}}

\author{Lech Szymanowski\\
        National Center for Nuclear Research (NCBJ), Warsaw, Poland\\
       E-mail: \email{Lech.Szymanowski@fuw.edu.pl}}
				
\author{Samuel Wallon\\
       LPT, Universit\'e Paris-Sud, CNRS, 91405, Orsay, France\\
       UPMC Univ. Paris 06, facult\'e de physique, 4 place Jussieu, 75252 Paris Cedex 05,
France\\
       E-mail: \email{Samuel.Wallon@th.u-psud.fr}}

\abstract{We present a phenomenological model for the helicity amplitudes $T_{11}$ and $T_{00}$ of the rho meson exclusive diffractive leptoproduction in the forward limit. This model leads to a very good description of the polarized cross-sections $\sigma_T$ and $\sigma_L$ when compared to HERA data. This model is based on the impact factor representation of the helicity amplitudes. The $\gamma^*\to\rho$ impact factor is computed within the light-cone collinear factorization scheme, the impact parameter space representation allowing to factorize out the dipole-target amplitude. Finally our description combines a model for the dipole-target amplitude that includes the saturation effects with the results for the impact factor where the twist~2 and twist~3 distribution amplitudes of the rho meson are involved.}

\FullConference{XXI International Workshop on Deep-Inelastic Scattering and Related Subject -DIS2013,\\
		22-26 April 2013\\
		Marseille, France}

\newcommand{\kb}{\underline{k}}
\newcommand{\garho}{\gamma^*_{\lambda_{\gamma}}\to \rho_{\lambda_{\rho}}}
\newcommand{\garhoL}{\gamma^*_{L}\to \rho_{L}}
\newcommand{\garhoT}{\gamma^*_{T}\to \rho_{T}}
\newcommand{\qb}{\bar{q}}
\newcommand{\rb}{\underline{r}}
\newcommand{\yb}{\bar{y}}
\newcommand{\nn}{\nonumber}
\begin{document}

\section{Introduction}
\begin{figure}[htbp]
	\centering
\psfrag{P}[cc][cc]{$p$}
\psfrag{P'}[cc][cc]{$p'$}
\psfrag{Gam}[cc][cc]{$\gamma^*$}
\psfrag{Rho}[cc][cc]{$\rho$}
\psfrag{k}[cc][cc]{$k_1$}
\psfrag{k2}[cc][cc]{$k_2$}
\psfrag{Phi}[cc][cc]{${\Phi^{\gamma^*\to\rho}}$}
\psfrag{Phibis}[cc][cc]{$\mathcal{F}(x,\kb)$}
\includegraphics[width=0.25\linewidth]{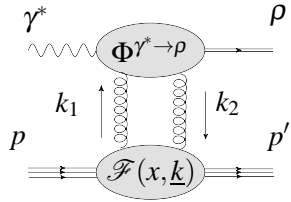}
\caption{Impact factor representation of the helicity amplitudes.}
	\label{fig:ImpFactDVMP}
\end{figure}

We present a phenomenological model for the longitudinal and transverse polarized cross-sections of the exclusive diffractive leptoproduction of the rho meson in the high energy limit. The polarized cross-sections are obtained from the helicity amplitudes $T_{00}$ and $T_{11}$ in the forward limit $t \to 0$ where we denote $T_{\lambda_{\rho}\lambda_{\gamma}}$ the amplitude associated to the process
\begin{equation}
\label{process}
\gamma^*(\lambda_{\gamma},q)\,N(p)\to \rho(\lambda_{\rho},p_{\rho}) \, N(p')
\end{equation}
for a nucleon target $N$, and where $\lambda_{\gamma}$ and $\lambda_{\rho}$ denote the polarizations of the virtual photon and of the rho meson. Our approach is based on the following kinematical assumptions:
\begin{itemize}
\item the center of mass $\gamma^*\,N$ energy is asymptotically large and one can define two light-cone momenta $p_1$ and $p_2$ such that,
\begin{equation}
p_{\rho}\sim p_1\,,\quad p\sim p_2 \,, \quad q\sim p_1- \frac{Q^2}{s}p_2\,, \quad s=(q+p)^2\sim 2 p_1\cdot p_2\gg Q^2\,,\,\,m_{\rho}^2\,,
\end{equation}
\item the virtuality of the photon $Q$ is much larger than the QCD scale $\Lambda_{QCD}$ in order to compute the photon vertex using pQCD techniques. 
\end{itemize}
The first assumption allows to factorize helicity amplitudes $T_{\lambda_{\rho}\lambda_{\gamma}}$ into the $\garho$ impact factor
\begin{equation}
\label{impactdef}
\Phi^{\gamma^*_{\lambda_{\gamma}}\to\rho_{\lambda_{\rho}}}=\frac{1}{2s} \int \frac{d \kappa}{2\pi} i \mathcal{M}(\gamma^*(\lambda_{\gamma},q)+g(k_1)\to \rho(\lambda_{\rho},p_1)+g(k_2))\,,
\end{equation}
where $\mathcal{M}$ is the amplitude of the sub-process $\gamma^*(\lambda_{\gamma},q)+g(k_1)\to \rho(\lambda_{\rho},p_1)+g(k_2)$, $\mathcal{F}(x,\kb)$ is the unintegrated gluon density and $\kappa=(q+k_1)^2$. The helicity amplitude $T_{\lambda_{\rho}\lambda_{\gamma}}$ then reads\footnote{We denote $\underline{x}$ the 2-dimension euclidean vector associated to the Minkowskian space vector $x_{\perp}$ in the transverse plane, $\underline{x}^2=-x_{\perp}^2$.} 
\begin{equation}
\label{ConvImp}
T_{\lambda_{\rho}\lambda_{\gamma}}= is \int \frac{d^2\kb}{(\kb^2)^2} \, \Phi^{\gamma^*_{\lambda_{\gamma}}\to\rho_{\lambda_{\rho}}}(\kb) \, \mathcal{F}(x,\kb)\,,
\end{equation}
 where $k_1=\frac{\kappa+Q^2+\kb^2}{s}p_2+k_{\perp}$ and $k_2=\frac{\kappa+\kb^2}{s}p_2+k_{\perp}$ are the $t-$channel gluon momenta 
(fig.~\ref{fig:ImpFactDVMP}).

The second assumption $Q^2\gg \Lambda^2_{QCD}$ allows to compute the $\garho$ impact factor within the light-cone collinear factorization scheme. The leading twist $\garhoL$ impact factor has been computed in \cite{GinzburgPanfilSerbo} by Ginzburg, Panfil and Serbo in 1987 while an approach to derive the $\garhoT$ impact factor, which involves the twist~3 light-cone operators of the rho meson, have been performed in 2010 in~\cite{Anikin2009,Anikin2010} by Anikin, Ivanov, Pire, Szymanowski and Wallon. 
The results obtained from this first principle approach are parameterized by the rho meson twist~2 and twist~3 distribution amplitudes (DAs). Using the model 
 of ref.~\cite{Ball:1998sk} from Ball, Braun, Koike and Tanaka for the DAs, 
 we built a first model in ref.~\cite{Anikin2011}, using 
a phenomenological model for the nucleon impact factor that was proposed in ref.~\cite{GunionSoper} by Gunion and Soper in 1977. 
The sizable contribution from soft gluons of $\kb^2<1\;$GeV$^2$ motivates the present model where saturation effects are taken into account.

\begin{figure}[htbp]
	\centering
	\begin{tabular}{c}
	\psfrag{Hti}{$\tilde{H}^{\Gamma}_{q\qb}(y,\rb)$}
	\psfrag{S}{$S_{q\qb}^{\Gamma}(y)$}
	\psfrag{r}{$\rb$}
	\psfrag{y}{$y$}
	\psfrag{yb}{$\yb$}
	\psfrag{rS}{$r_{\perp} \cdot S^{\Gamma}_{q\qb\perp}(y)$}
	\psfrag{Pl}{$+$}
	\psfrag{Plc}{$+\cdots$}
	\psfrag{INT}{$\int dy\int d^2\rb$}
		\includegraphics[width=0.76\textwidth]{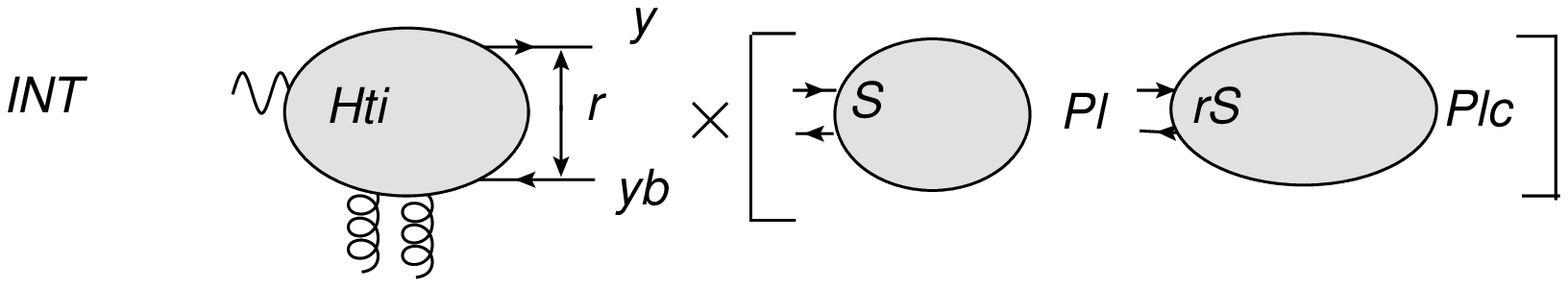}\\ 
		\Large{+}\\
	\psfrag{Hti}{$\tilde{H}^{\Gamma}_{q\qb g}(y_i,\rb_i)$}
	\psfrag{S}{
	$S^{\Gamma}_{q\qb g}(y_1,y_2)$}
	\psfrag{r}{$\rb_1$}
	\psfrag{rb}{$\rb_2$}
	\psfrag{y}{$y_1$}
	\psfrag{yb}{}
	\psfrag{yg}{$y_g$}
	\psfrag{Plc}{$+\cdots$}
	\psfrag{INT}{\hspace{-1cm} $\int dy_1dy_g\int d^2\rb_1\int d^2\rb_2$}
	\includegraphics[width=0.78\textwidth]{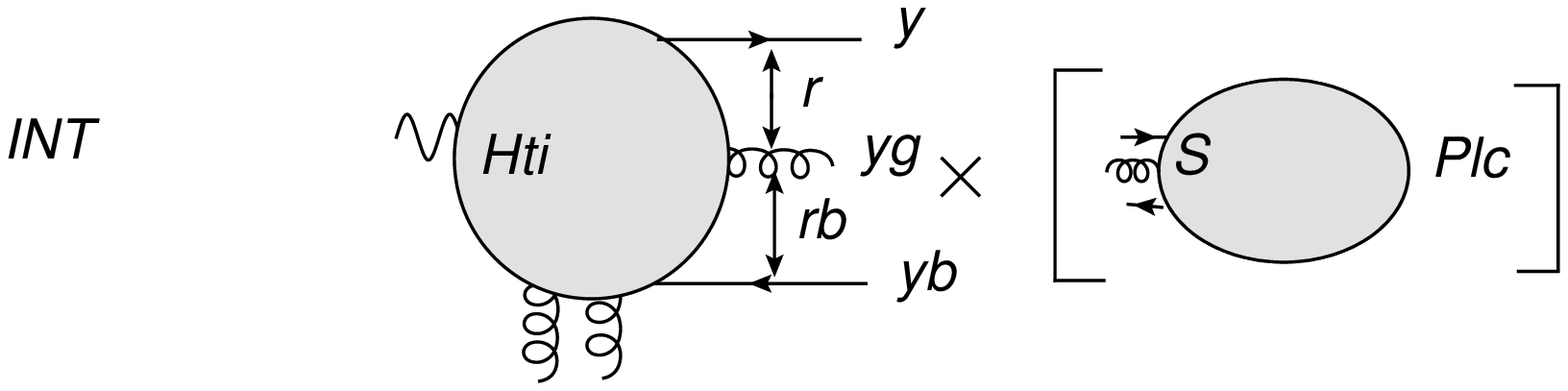}
	\end{tabular}
	\caption{Contributions to the twist~3 impact factor. The $\tilde{H}^{\Gamma}$'s are Fourier transforms in the transverse coordinate space of hard parts and the $S^{\Gamma}$'s functions are soft parts that are parameterized by the DAs up to twist~3. The $y$'s stand for the longitudinal fractions of rho meson momentum of the partons and $\rb$'s stand for the transverse sizes between the partons. 
	The label $\Gamma$ indicates the Fierz structure on which are projected the different contributions in order to factorize hard and soft parts in spinor space.}
	\label{TaylorFourier}
\end{figure}

In ref.~\cite{Besse2012} we have shown that the impact factor in the impact parameter representation reads,
\begin{eqnarray}
\label{phiLpsi}
\Phi^{\gamma^{*}_L \rightarrow \rho_L}(\kb,Q,\mu^2)&=&\left(\frac{\delta^{ab}}{2}\right)\int dy \int d \rb\,\, \psi^{\gamma^*_L\to\rho_L}_{(q\qb)}(y,\rb;Q,\mu^2)\, \mathcal{A}(\rb,\kb)\,,\\
\Phi^{\gamma^{*}_T \rightarrow \rho_T}(\kb,Q,\mu^2)&=&\left(\frac{\delta^{ab}}{2}\right)\int dy \int d \rb\,\, \psi^{\gamma^*_T\to\rho_T}_{(q\qb)}(y,\rb;Q,\mu^2) \,\mathcal{A}(\rb,\kb)\nn\\
&&\hspace{-1cm}+\left(\frac{\delta^{ab}}{2}\right)\int dy_2 \int dy_1 \int d \rb \,\, \psi^{\gamma^*_T\to\rho_T}_{(q\qb g)}(y_1,y_2,\rb;Q,\mu^2) \mathcal{A}(\rb,\kb)\,,
\label{phiTpsi}
\end{eqnarray}
where the functions $\psi^{\gamma^*_{L(T)}\to \rho_{L(T)}}$ are the results for the overlaps of the wave functions of the virtual photon and the rho meson, computed in the collinear factorization approach up to twist~3 and $\mathcal{A}$ is the dipole-target amplitude. The computation of the $\garhoT$ transition involves the quark antiquark contribution and the quark antiquark gluon contribution as represented in the fig.~\ref{TaylorFourier}.

Soft parts $S^{\Gamma}_{q\qb}$ and $S^{\Gamma}_{q\qb g}$ are parameterized by a set of six twist~3 and one twist~2 DAs which are not independent. This set can be reduced to a set of three independent DAs for which the model in ref.~\cite{Ball:1998sk} gives explicit expressions depending on the renormalization scale $\mu^2$. The result in the limit $\mu^2\to\infty$ is called "asymptotic" (AS) result. The full twist~3 result (Total), where we put $\mu^2=\frac{Q^2+m_{\rho}^2}{4}$, can be also separated into two contributions, the Wandzura-Wilczek (WW) contribution that only depends on the twist~2 DA and the genuine contribution which only depends on the quark antiquark gluon (twist~3) DAs.

The factorization of the dipole-target scattering amplitude in eqs.~(\ref{phiLpsi}, \ref{phiTpsi}) 
 allows to implement arbitrary models for this dipole-target amplitude. 
 Neglecting the skewness effects in the dipole-target scattering amplitude, the helicity amplitudes can be expressed in terms of the dipole cross-section $\hat{\sigma}(x,r)$,
\begin{eqnarray}
\label{T00Lpsi}
\frac{T_{00}}{s}\!&\!=\!&\!\int dy \int d \rb\, \psi^{\gamma^*_L\to\rho_L}_{(q\qb)}(y,\rb;Q,\mu^2) \,\hat{\sigma}(x,\rb)\,,\\
\label{T11Tpsi}\frac{T_{11}}{s}\!&\!=\!&\! \int d \rb\left[\int \! dy \psi^{\gamma^*_T\to\rho_T}_{(q\qb)}(y,\rb;Q,\mu^2)
+\int \! dy_2 \int \! dy_1  \psi^{\gamma^*_T\to\rho_T}_{(q\qb g)}(y_1,y_2,\rb;Q,\mu^2)\right] \hat{\sigma}(x,\rb)\,.
\end{eqnarray}
Assuming the phenomenological $t-$dependence of the differential cross-sections,
 \begin{equation}
\label{t-dependence}
\frac{d \sigma_{L,T}}{d t}(t)=e^{-b(Q^2) t}\,\frac{d \sigma_{L,T}}{d t}(t=0)\,,
\end{equation}
where $b(Q^2)$ has been extracted from the H1 data \cite{H1}, the polarized cross-sections $\sigma_L$ and $\sigma_T$ can be expressed in terms of the forward helicity amplitudes $T_{00}$ and $T_{11}$,
\begin{equation}
\label{sigmaLprediction}
\sigma_L =\frac{1}{b(Q^2)}\frac{\left|T_{00} (s,t=0)\right|^2 }{16\pi s^2}\,,\quad
\sigma_T =\frac{1}{b(Q^2)}\frac{\left|T_{11} (s,t=0)\right|^2 }{16\pi s^2}\,.
\end{equation}
\section{Results}
\begin{figure}[htbp]
	\centering
	\begin{tabular}{cc}
		\includegraphics[width=0.450\textwidth]{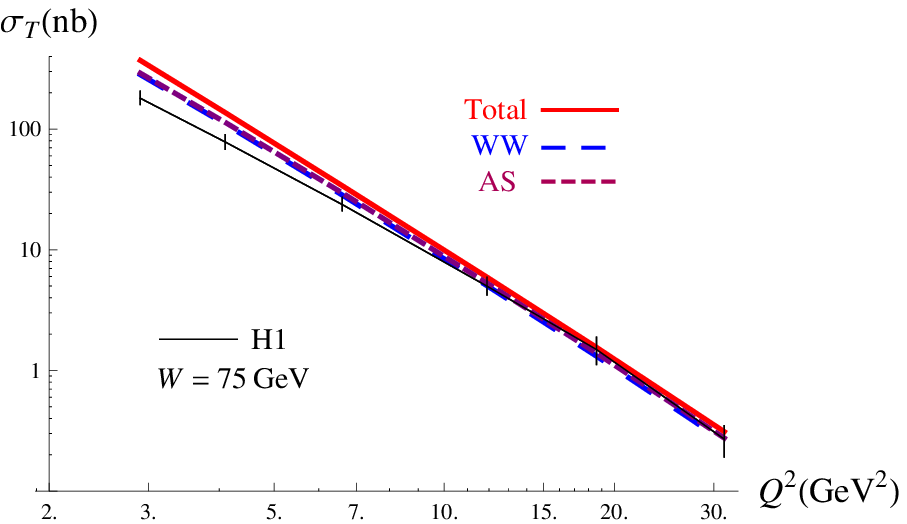}&\includegraphics[width=0.45\textwidth]{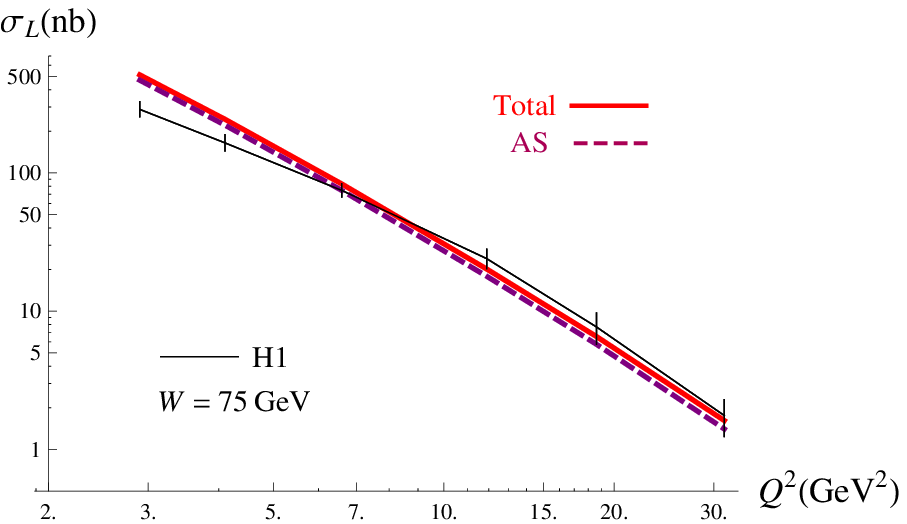}\\
	\includegraphics[width=0.45\textwidth]{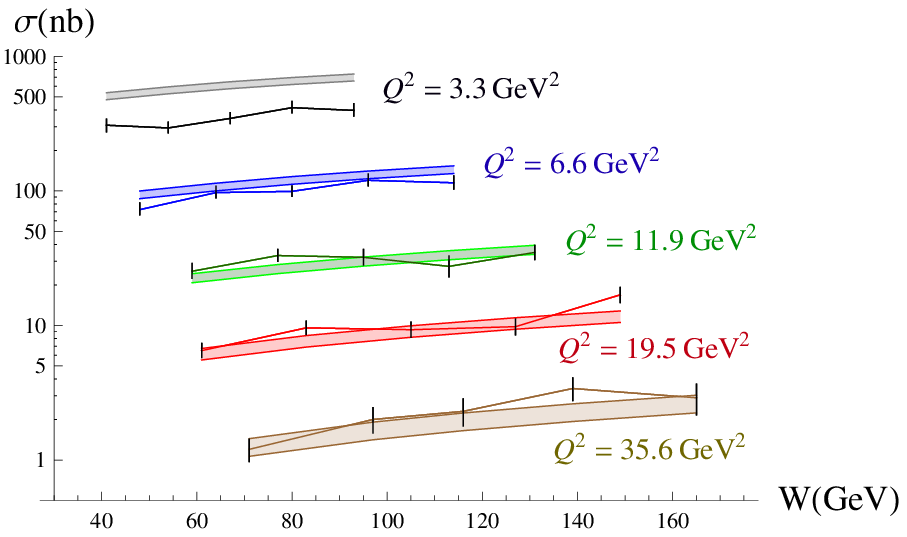}	& \includegraphics[width=0.45\textwidth]{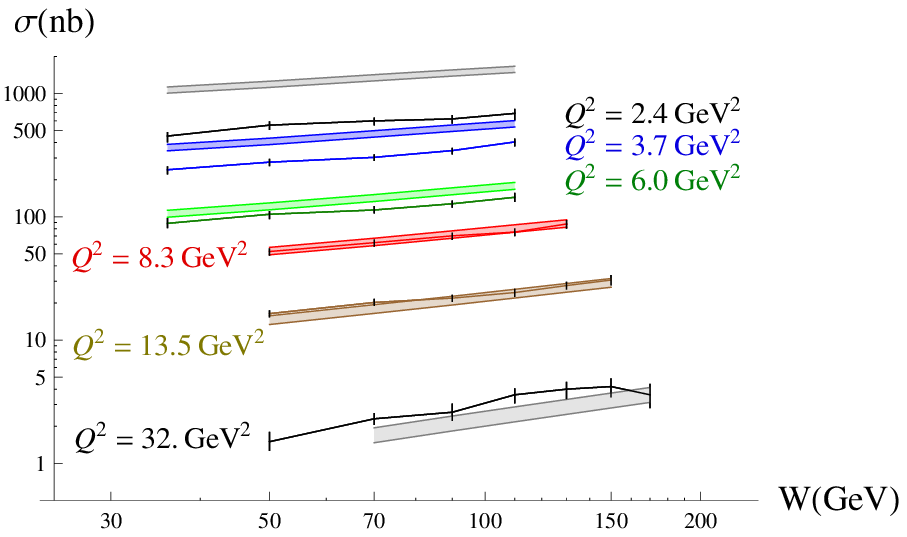}
	\end{tabular}
	\caption{Top left: Total, WW and AS contributions to $\sigma_T$ vs $Q^2$, compared to H1 \cite{H1} data. Top right: Total and AS twist~2 contributions to $\sigma_L$ vs $Q^2$ compared to H1 data. Bottom line: Predictions for the total cross-section $\sigma$ vs $W$ compared to H1 (left) and ZEUS \cite{ZEUS} (right) data.}
	\label{Fig3}
\end{figure}

In figs.~\ref{Fig3} are shown the predictions of our model, from ref.~\cite{Besse:2013muy}, obtained using the dipole cross-section model of ref.~\cite{Albacete:2010sy}, compared with H1 \cite{H1} and ZEUS \cite{ZEUS} data. 
For $Q^2\gtrsim 5\;$GeV$^2$, the predictions we obtain, without any free parameter to adjust, are in very good agreement with the $Q^2-$ and the $W-$dependence of the polarized cross-sections. The success of this model to describe these dependence and the normalizations of the cross-sections indicates that the factorization scheme which is chosen in this study works for large $Q^2$. The discrepancy for small $Q^2\lesssim 5\;$GeV$^2$ could be due to higher twist corrections of the $\garho$ impact factors. Note that 
the cross-sections have a weak dependence in the choice of the renormalization scale $\mu$.

\section{Conclusion}

 We have presented a model based on first principle calculations to factorize the helicity amplitudes. The non-perturbative parts of the process are encoded in the DAs of the rho meson and the dipole scattering amplitude. We use the universality of these non-pertubative objects to get a model without free parameter.
The predictions obtained for the polarized cross-sections of the rho meson diffractive leptoproduction works successfully for large $Q^2$. Higher twist corrections would be desirable in order to get a better control for lower $Q^2$ values and thus to get closer to the genuine saturation regime in the HERA kinematics. The extension of this treatment in the non-forward limit would allow to get the dipole-target impact parameter dependence, which would be a good probe to determine the gluon density profile of the proton in the high energy limit.
\section{Acknowledgment}
We thank B. Duclou\'e, K. Golec-Biernat, C. Marquet, S. Munier and B. Pire for interesting discussions and comments on this work. This work is supported in part by the french grant ANR PARTONS (ANR-12-MONU-0008-01), the Polish Grant NCN No DEC-2011/01/D/ST2/03915 and the Joint Research Activity Study of Strongly Interacting Matter (acronym HadronPhysics3, Grant 283286) under the Seventh Framework Programme of the European Community.

\end{document}